\documentclass[english,floatfix,showkeys,superscriptaddress,nofootinbib,aps,prd]{revtex4}
\usepackage[utf8]{inputenc}
\usepackage[T1]{fontenc}
\usepackage{lmodern}
\usepackage{amsmath}
\usepackage{amssymb}
\usepackage{graphicx}
\usepackage{esint}
\usepackage{longtable}
\usepackage{dcolumn}
\usepackage{babel} 
\usepackage{csquotes} 
\usepackage{color} 

\begin{document}

\title{Viable wormhole solution in Bopp--Podolsky electrodynamics}

\author{D. A. Frizo}
\email{diego.frizo@ufabc.edu.br}
\affiliation{Centro de Matemática, Computação e Cognição, Universidade Federal do ABC, \\  
Avenida dos Estados 5001, CEP 09210-580, Santo André, SP, Brazil}

\author{C. A. M. de Melo}
\email{cassius.melo@unifal-mg.edu.br}
\affiliation{Instituto de Ciência e Tecnologia, Universidade Federal de Alfenas, \\ Rodovia José Aurélio Vilela,
11999, CEP 37715-400 Poços de Caldas, MG, Brazil}

\author{L. G. Medeiros}
\email{leogmedeiros@gmail.com}
\affiliation{Escola de Ciência e Tecnologia, Universidade Federal do Rio Grande do Norte, Campus Universitário, s/n-Lagoa Nova, Natal, RNCEP 59078-970, Brazil} 

\author{Juliano C. S. Neves}
\email{juliano.c.s.neves@gmail.com}
\affiliation{Instituto de Ciência e Tecnologia, Universidade Federal de Alfenas, \\ Rodovia José Aurélio Vilela,
11999, CEP 37715-400 Poços de Caldas, MG, Brazil}


\begin{abstract}
Following a recent approach in which the gravitational field equations in curved spacetimes were 
presented in the Bopp--Podolsky electrodynamics, 
we obtained an approximate and spherically symmetric wormhole solution in this context.
The calculations were carried out up to the linear approximation in both the spacetime geometry 
and the radial electric field. The solution presents a new parameter 
 that comes from the Lagrangian of the model. Such a parameter was constrained
 by using the shadow radius of Sagittarius A*, 
 recently revealed by the Event Horizon Telescope Collaboration. Remarkably, 
the wormhole presented here is viable when its shadow is compared to the Sagittarius A* shadow.
\end{abstract}


\maketitle

\section{Introduction}
The Bopp--Podolsky electrodynamics \cite{Bopp,Podolsky} is an extension 
of Maxwell's theory in which a new second-order
derivative of the gauge field is present in the Lagrangian, leading then to high-order field equations with fourth-order 
terms. As is well known, high-order theories in this context contain instabilities (like ghost instabilities).
However, the Bopp--Podolsky electrodynamics is able to avoid this situation by means of the concept of Lagrange
anchor \cite{Kaparulin:2014vpa}. 
Among the reasons to work on the Bopp--Podolsky model is its capability to preserve
the linearity of the field equations being the only second-order gauge theory for the $U(1)$ group that
accomplishes that \cite{Cuzinatto:2005zr}. Moreover, in $3+1$ dimensions (flat spacetime), the Bopp--Podolsky 
term in the Lagrangian can also be conceived of as an effective term coming from quantum 
corrections to the photon action in the low-energy regime \cite{Borges:2019gpz}.

Specific studies on the Bopp--Podolsky context have been done focusing, for example,
 on self-interaction \cite{Zayats:2013ioa,Zayats:2016iun}, dark energy \cite{Haghani:2016oxv}, 
 massive photons in cosmology \cite{Cuzinatto:2016kjk} and spherically symmetric black holes \cite{Cuzinatto}. 
 In this work,  we built an approximate solution with spherical
symmetry extending the results of Ref. \cite{Cuzinatto}.
But contrary to the mentioned article \cite{Cuzinatto}, in which the spacetime geometry 
was studied without an analytic expression
for the geometry, here we present a first-order approximation in order to obtain an explicit wormhole metric in
Bopp--Podolsky electrodynamics.
Such as in Cuzinatto et al. \cite{Cuzinatto}, our approximate wormhole solution also violates both the null (NEC)
 and the weak energy (WEC) conditions (energy conditions 
 violations are important consequences in research areas like regular black holes \cite{Neves:2014aba} and 
 bouncing cosmologies \cite{Novello:2008ra,Brandenberger:2016vhg} in today's physics). 
 In wormhole physics,
 violations of energy conditions at the wormhole throat are commonly reported \cite{Morris:1988cz,Visser:2003yf,Lemos:2003jb,Bronnikov:2002rn,Lobo:2005us,Molina,Molina:2012ay,Muniz:2022eex}.
 Like the spacetime metric in the mentioned work \cite{Cuzinatto}, our solution also 
brings out the coupling constant $b$. In the mentioned article, because of $b$,
the authors say their solution would not be in agreement with the no-hair conjecture.

The no-hair conjecture/theorem \cite{Israel:1967wq,Israel:1967za,Carter:1971zc} 
states that black holes are fully 
described by just three parameters: mass, charge and spin. Contrary to Cuzinatto et al. \cite{Cuzinatto}, 
we do not interpret the new parameter $b$
as some sort of hair. Besides, we claim that the solution of Ref. \cite{Cuzinatto}
is not a black hole, but a wormhole solution.
Even so, a useful task is indicating constraints on the new parameter $b$
unfolded by our approximate metric.

In this regard, we compare the shadow angular diameter or radius of the spacetime metric obtained here 
with the recent image of the Sagittarius A* (Sgr A*) shadow released by the Event Horizon Telescope (EHT) \cite{EventHorizonTelescope:2022xnr,EventHorizonTelescope:2022xqj}. 
Black hole and wormhole shadows have been a seminal research area since 
the very first shadow revealed by the EHT, the M87* shadow
\cite{EventHorizonTelescope:2019dse,EventHorizonTelescope:2019ggy}.
Two shadow parameters for constraining general relativity and modified theories of gravity have been
adopted recently in the literature, 
namely the shadow angular diameter $d_{\text{sh}}$ \cite{Maluf:2020kgf,Khodadi:2020jij,Vagnozzi,Khodadi:2022pqh,KumarWalia:2022aop,Afrin:2022ztr}
 and the shadow deviation 
from circularity $\Delta C$ \cite{Bambi:2019tjh,Kumar:2020ltt,Neves:2019lio,Neves:2020doc,Banerjee:2019nnj,Khodadi:2021gbc,Kumar:2020yem}, 
which is nonzero for rotating black holes and is zero for nonrotating ones, that is
to say, the shadow of spherical black holes or wormholes is a perfect circle.
In particular, for the Sgr A* black hole (or wormhole), due to the more precise measurements of 
distance and mass, the EHT \cite{EventHorizonTelescope:2022xqj} used the shadow angular 
diameter in order to provide a new parameter to be compared to black hole or wormhole metric candidates, 
whether in the general relativity
context or in modified theories of gravity. From the shadow angular diameter, the new parameter 
called deviation from the Schwarzschild metric ($\delta$) arises from the black hole 
mass-to-distance ratio  and has been adopted, for example, in the pretty complete review on alternative
theories of gravity and spherically symmetric black hole and wormhole metrics carried out 
by Vagnozzi et al. \cite{Vagnozzi}. 
According to the authors of the mentioned review, 
the deviation from the Schwarzschild metric is, for example, a very good parameter to constraint spherical and
 \enquote{hairy} black holes, and wormholes by using the shadow of Sgr A*.
That is the reason why we are going to follow the approach presented in Ref. \cite{Vagnozzi} in order
to indicate that the metric obtained here is viable, i.e., for a range of parameters (mass, charge and 
coupling constant of the model), the spherically symmetric wormhole in the
Bopp--Podolsky electrodynamics with the same mass of Sgr A*
 is able to produce a shadow similar to that one captured by the EHT.

This article is structured as follows: in Sec. II the action, the electric field equations (also called Podolsky
equations) and the gravitational 
 field equations are presented for the Bopp--Podolsky
electrodynamics in curved spacetimes. 
Both the gravitational and the electric field equations are solved in the linear approximation in Sec. III.
Sec. IV is an attempt at indicating that the approximate solution is viable from the observational point of view
according to the Sgr A* shadow. The final comments are in Sec. V.
In this article, we are adopting the geometrized units, that is, $G=c=1$, where $G$ is the gravitational constant, and
$c$ is the speed of light in vacuum.

\section{The Bopp--Podolsky electrodynamics for curved spacetimes}
\label{S2}
In this section, both the action and field equations for curved spacetimes in the Bopp--Podolsky
electrodynamics are written and briefly commented. 
According to Cuzinatto et al. \cite{Cuzinatto}, the Lagrangian in curved spacetimes for the 
Bopp--Podolsky electrodynamics is described 
by two new independent and invariant terms beyond the usual Maxwell term. The Lagrangian in this context
is: (i) invariant under Lorentz transformations, (ii) gauge invariant under the $U(1)$ symmetry group, (iii) 
quadratic in the gauge field and its derivatives, and (iv) dependent on the gauge field and its first two derivatives.
As will see, the two new invariant terms in the Lagrangian are either minimally or nonminimally coupled 
to gravity. Assuming then this four conditions,  
the Bopp--Podolsky electrodynamics for curved spacetimes is described by the following Lagrangian:
\begin{equation}
\mathcal{L}_{m} =- \frac{1}{4}F^{\alpha\beta}F_{\alpha\beta}+\frac{\left(
a^{2}+2b^{2}\right)  }{2}\nabla_{\beta}F^{\alpha\beta}\nabla_{\gamma}%
F_{\alpha}^{\text{ \ }\gamma}
+b^{2}\left(  R_{\sigma\beta}F_{\text{ \ }}^{\sigma\alpha}F_{\alpha}^{\text{
\ }\beta}+R_{\alpha\sigma\beta\gamma}F^{\sigma\gamma}F^{\alpha\beta}\right),
\label{Lagrangian}
\end{equation}
where $F_{\mu\nu}=\nabla_{\mu}A_{\nu}-\nabla_{\nu}A_{\mu}$ is the field strength, $\nabla_{\mu}$ is
the covariant derivative,  $a$ and $b$ are coupling constants, $R_{\mu\nu}$ is the Ricci tensor, and
$R_{\alpha\beta\gamma \delta}$ is the Riemann tensor.

By using the so-called Einstein--Hilbert action plus the term coming from the Bopp--Podolsky electrodynamics, namely
\begin{equation}
S=\frac{1}{16\pi}\int d^4x \sqrt{-g}\left(-R+4 \mathcal{L}_m \right),
\label{Action}
\end{equation}
in which $g$ is the metric determinant, $R$ is the Ricci scalar, 
and $\mathcal{L}_{m}$ is given by Eq. (\ref{Lagrangian}), we are
able to obtain the gravitational field equations and the equations for the gauge field, also called 
Bopp--Podolsky equations.
In this regard, by varying  the action (\ref{Action}) with respect to the metric field $g_{\mu\nu}$, one has
the corresponding gravitational field equations: 
\begin{equation}
R_{\mu\nu}-\frac{1}{2}g_{\mu\nu}R= 8\pi\left(  T_{\mu\nu}^{M}+T_{\mu\nu}^{a}+T_{\mu\nu}^{b}\right),
\label{Eq_gravitacional}%
\end{equation}
where the components of the energy-momentum tensor are given by
\begin{eqnarray}
T_{\mu\nu}^{M} &=&\frac{1}{4\pi}\left[  F_{\mu\sigma}F_{\text{ \ }\nu
}^{\sigma}+g_{\mu\nu}\frac{1}{4}F^{\alpha\beta}F_{\alpha\beta}\right],
\label{T_munu M}\\
T_{\mu\nu}^{a} &=&\frac{a^{2}}{4\pi}\left[  g_{\mu\nu}F_{\beta}^{\;\gamma
}\nabla_{\gamma}K^{\beta}+\frac{g_{\mu\nu}}{2}K^{\beta}K_{\beta}\right.
 +\left.  2F_{(\mu}^{~\alpha}\nabla_{\nu)}K_{\alpha}-2F_{(\mu}^{~\alpha
}\nabla_{\alpha}K_{\nu)}-K_{\mu}K_{\nu}\right], \label{T_munu a}\\
T_{\mu\nu}^{b} &=&\frac{b^{2}}{2\pi}\left[  -\frac{1}{4}g_{\mu\nu}%
\nabla^{\beta}F^{\alpha\gamma}\nabla_{\beta}F_{\alpha\gamma}+F_{\text{ }(\mu
}^{\gamma}\nabla^{\beta}\nabla_{\beta}F_{\nu)\gamma}\right. 
+ \left.  F_{\gamma(\mu}\nabla_{\beta}\nabla_{\nu)}F^{\beta\gamma}%
-\nabla_{\beta}\left(  F_{\gamma}^{\text{ \ }\beta}\nabla_{(\mu}F_{\nu
)}^{\text{ \ }\gamma}\right)  \right].  \label{T_munu b}%
\end{eqnarray}
The notation $\left(  ...\right)  $ indicates symmetrization with respect to the indexes inside the brackets.

On the other hand, by varying the action (\ref{Action}) with respect to the field $A_{\mu}$, 
we have the Bopp--Podolsky equations for curved spacetimes, namely
\begin{equation}
\nabla_{\nu}\left[  F^{\mu\nu}-\left(  a^{2}+2b^{2}\right)  H^{\mu\nu}%
+2b^{2}S^{\mu\nu}\right]  =0,\label{Eq_Podolsky}
\end{equation}
where
\begin{eqnarray}
H^{\mu\nu} &\equiv &\nabla^{\mu}K^{\nu}-\nabla^{\nu}K^{\mu},\label{H_munu}\\
S^{\mu\nu} &\equiv &F_{\text{ \ }}^{\mu\sigma}R_{\sigma}^{\text{ \ }\nu
}-F^{\nu\sigma}R_{\sigma}^{\text{ \ }\mu}+2R_{\text{ \ }\sigma\text{ }\beta
}^{\mu\text{ \ }\nu}F^{\beta\sigma},\label{S_munu}%
\end{eqnarray}
with $
K^{\mu}\equiv\nabla_{\gamma}F^{\mu\gamma}$. It is worth pointing out that any 
valid solution in the Bopp--Podolsky
 context must be solution of either field equations, whether the gravitational field equations (\ref{Eq_gravitacional}) or 
the Bopp--Podolsky equations (\ref{Eq_Podolsky}).

In Ref. \cite{Cuzinatto}, by using the Bekenstein method, even without an explicit form for the spacetime metric,
it was shown that for $b=0$ the exterior solution of a spherically symmetric black hole 
is the Reissner--Nordstr\"{o}m solution, i.e., the no-hair theorem is fully satisfied. 
However, the same authors claimed that hairy black hole solutions would be possible only 
if both $b\neq0$ and $\frac{dg_{00}}{dr}\geq 0$, where $r$ is the radial coordinate. 
In the next section, using a perturbative approach, we 
then show explicitly that the $b\neq0$ case generates instead a wormhole solution.

From the energy conservation perspective, the total energy-momentum tensor is conserved, i.e.,
given $T_{\mu\nu}=T_{\mu\nu}^M+T_{\mu\nu}^a+T_{\mu\nu}^b$, one has
\begin{equation}
\nabla_{\nu}T^{\mu\nu}=0,
\label{Conservation}
\end{equation} 
with the aid of the constraint provided by the Bopp--Podolsky equations (\ref{Eq_Podolsky}). 
In particular, for the spherically symmetric \textit{Ansatz} that we used in this study
(independently of the metric terms) and a radial electric field, every component of 
Eq. (\ref{Conservation}) is identically zero, regardless the $\mu=1$ component. Such a component
becomes zero by using the $\mu=1$ component of the Bopp--Podolsky equations.

\section{An approximate solution}
\label{Solution}

\subsection{The spacetime metric}
When dealing with higher-order theories in which the coupling parameters are very small, 
the correct perturbative scheme is given by the singular perturbation theory \cite{Johnson,Holmes}. 
The influence of higher derivatives is important inside a boundary layer, whose size is proportional 
to the coupling constants coming from the higher-order derivatives, while outside that boundary 
the solution is asymptotically the nonperturbed solution, and the regular approach for obtaining approximate
solutions is enough. In our case, the higher-order derivatives of the electric field 
are proportional to the parameters or coupling constants $a$ and $b$. 
Since we are interested in the exterior solution, we are dealing with scale distances such that 
(in geometrized units)
\begin{equation}
a, b \ll r_+ \leq r< \infty,
\end{equation}
where $r_+$ is the radius of the external horizon, the usual event horizon radius, in a black hole solution. 
This implies that we are able to use only the regular perturbation theory throughout this article.

Having said that, we present here an approximate solution with spherical symmetry 
in the context indicated in Section \ref{S2}
by using the regular perturbation theory.
Due to the nonlinearity of the field equations (as for the metric components), 
approximate solutions are an alternative alongside the numerical
approach. From the spherically symmetric \textit{Ansatz} in the $(t,r,\theta,\phi)$ coordinates, i.e.,
\begin{equation}
ds^2 = A(r)dt^2-\frac{dr^2}{B(r)}-r^2 \left(d\theta^2 +\sin^2 \theta d\phi^2 \right),
\label{SS}
\end{equation}  
an approximate solution will be available if we assume
\begin{align}
A(r) & = A_0 (r)+\epsilon A_1 (r), \label{A} \\
B(r) & = B_0 (r) + \epsilon B_1 (r), \label{B}  \\
E(r) & = E_0 (r) + \epsilon E_1 (r). \label{E}
\end{align}
As we are looking for a small deviation from the Reissner--Nordström geometry, then we assume the following
input for the metric and the radial electric field, i.e.,
\begin{equation}
A_0 (r) = B_0 (r)= 1- \frac{2M}{r}+\frac{Q^2}{r^2} \hspace{0.3cm} \mbox{and} \hspace{0.3cm} E_0 (r)=\frac{Q}{r^2},
\label{A0E0}
\end{equation}
where $\epsilon=\epsilon(a,b)$ is a small quantity,  and we take the perturbations just to the linear order.

Another important input of the model is the field strength $F_{\mu\nu}$.
From the small deviation of the Reissner--Nordström black hole in which we are interested, we assume that
\begin{equation}
F_{\mu\nu}=E(r) \left[ \delta^{1}_{\mu} \delta^{0}_{\nu}- \delta^{0}_{\mu} \delta^{1}_{\nu} \right].
\label{F}
\end{equation}  
Thus, the system of equations provided by the gravitational field equations depends on just three
functions: $A(r),B(r)$ and $E(r)$ or, in the perturbative approach that we adopt,
 $A_1(r),B_1(r)$ and $E_1(r)$.

The field equations (\ref{Eq_gravitacional}), by using the approximation  (\ref{A})-(\ref{E}) and the field strength
(\ref{F}), give us the following system of differential equations: 
\begin{align}
& B_1'(r) + \left( \frac{Q^2 +F(r)}{r F(r)}\right) B_1(r)-\frac{Q^2}{r F(r)} A_1(r) +\frac{2Q}{r}E_1(r) 
-\frac{4b^2Q^2}{\epsilon r^7} \Big(7Q^2 +\left(3r-10M \right)r \Big)=  0, \label{Equations1}\\
& A_1'(r) - \left( \frac{r^2 - F(r)}{r F(r)}\right) A_1(r) + \frac{r }{F(r)}B_1(r)+\frac{2Q}{r}E_1(r) 
- \frac{4b^2Q^2}{\epsilon r^7}\Big(Q^2 - \left(3r-2M \right)r \Big)=  0, \label{Equations2} \\
& A_1''(r) + \left(\frac{Q^2-Mr + F(r)}{r F(r)}\right) A_1'(r)+\left(\frac{r-M}{F(r)}\right) B_1'(r)  
  -\frac{4Q}{r^2}E_1(r) +\left(2\frac{\left(r-M \right)\left(Mr-Q^2 \right)}{r F(r)^2}\right) \big(A_1(r)-B_1(r) \big) \nonumber \\
& - \frac{16b^2Q^2}{\epsilon r^8} \Big(4Q^2+(3r-7M)r \Big) =  0,  
\label{Equations3}
\end{align}
with the operator $'$ playing the role of the derivative with respect to the radial coordinate, and 
\begin{equation}
F(r)= Q^2+\left(r-2M \right)r.
\end{equation}
An interesting point here is that by assuming the \textit{Ansatze} (\ref{A})-(\ref{A0E0}),
especially $E_0(r)$, all terms with $a$ disappear
in the field equations, because all terms coming from the invariant $\nabla_{\beta}F^{\alpha \beta}\nabla_{\gamma}F_{\alpha}^{\ \gamma}$ are considered zero in the approximation made here. 
That is, such an invariant term in the Lagrangian, owing to the mentioned \textit{Ansatze},
 brings the factor $(a^2+2b^2)\epsilon^2$, which is approximately zero. Therefore, only the nonminimal
terms will contribute with the perturbed equations, namely terms with $b^2$.

As we can see, we have three independent differential equations with three unknown 
functions: $A_1(r),B_1(r)$ and $E_1(r)$. 
In this regard, the system indicated in Eqs. (\ref{Equations1})-(\ref{Equations3}) is solvable,
and by assuming that the sough-after geometry is a small deviation from the Reissner--Nordström spacetime,
meaning that our solution is asymptotically flat, that is to say, 
the limits $\lim_{r\rightarrow \infty}A_1(r)=\lim_{r\rightarrow \infty}B_1(r)=\lim_{r\rightarrow \infty}E_1(r)=0$ are necessarily true, one has the following solution for the system
of differential equations:
\begin{align}
&A_1(r)=\frac{2b^2Q^2}{\epsilon r^4}\left(1-\frac{3M}{r}+\frac{9Q^2}{5r^2} \right), \label{A1}   \\
&B_1(r)=-\frac{2b^2Q^2}{\epsilon r^4}\left(2-\frac{3M}{r}+\frac{6Q^2}{5r^2} \right), \label{B1}  \\
&E_1(r)=-\frac{b^2Q}{\epsilon r^4}\left( \frac{8M}{r}-\frac{11Q^2}{r^2}\right). \label{E1}
\end{align} 
From the above result, we see that when Eqs. (\ref{A1})-(\ref{E1}) 
are substituted into Eq. (\ref{A})-(\ref{E}), 
the real expansion parameter will be $\frac{b}{r}\ll 1$ for a small deviation from the Reissner--Nordström spacetime. 
Then the parameter $\epsilon$ is some sort of auxiliary parameter
in the approach adopted here, something usual in standard procedures both in regular and
singular perturbation theory (see Refs. \cite{Johnson,Holmes}). 
It disappears once one substitutes Eqs. (\ref{A1})-(\ref{E1}) into Eqs. (\ref{A})-(\ref{E}).
In the end of the day,
the small deviation from the standard geometry will be given by the coupling constant $b$. Thus, $b$ is assumed
to be a very small parameter. 

It is worth pointing out that Eqs. (\ref{A1})-(\ref{E1}) are exact solutions of the system 
(\ref{Equations1})-(\ref{Equations3}), regardless the value of the radial coordinate $r$. 
Therefore, with Eqs. (\ref{A1})-(\ref{B1}) substituted into the 
linear approximation (\ref{A})-(\ref{B}), we arrive then in the fully calculated geometry, which
is written as
\begin{align}
ds^2=& \Bigg(1-\frac{2M}{r}+\frac{Q^2}{r^2} +\frac{2b^2Q^2}{r^4}-\frac{6Mb^2Q^2}{r^5}  
+\frac{18b^2Q^4}{5r^6} \Bigg) dt^2 -\Bigg(1-\frac{2M}{r}+\frac{Q^2}{r^2} 
 -\frac{4b^2Q^2}{r^4} +\frac{6Mb^2Q^2}{r^5}-\frac{12b^2Q^4}{5r^6} \Bigg)^{-1}dr^2 \nonumber \\
&-r^2\left(d\theta^2+\sin^2\theta d\phi^2 \right).
\label{Metric}
\end{align}
As is clear from the metric (\ref{Metric}), by making $b=0$ the Reissner--Nordström geometry is restored. And
contrary to the well-known black hole solutions with spherical symmetry in the general relativity context, 
the spacetime metric obtained here has $A(r)\neq B(r)$, which is something usual in modified theories of 
gravity.\footnote{In other contexts beyond general relativity, like braneworld,
the condition $A(r)\neq B(r)$ is quite common. See Refs. \cite{Molina,Molina:2012ay}.} 

As we can read from Eq. (\ref{Metric}), the metric 
presents a new parameter $b$, and this new parameter could be
interpreted as hair. Then the no-hair conjecture would be refuted. 
Cuzinatto et al. \cite{Cuzinatto} argue
in favor of a hairy black hole in the Bopp--Podolsky context. But, as we mentioned,
such authors argue without an explicit metric
like ours. In this regard, we disagree that $b$ would be a black hole parameter with the same status of mass, charge or
spin. By conceiving of $b$ as a constant of nature instead of a body property like mass or charge, 
the presence of $b$ is not in disagreement with the no-hair conjecture.
Following Cuzinatto et al. \cite{Cuzinatto}, the two conditions for a hairy and spherically 
symmetric black hole in Bopp--Podolsky electrodynamics, 
mentioned in Sec. \ref{S2}, namely $b\neq 0$ and $g'_{00}\geq 0$, are fulfilled
by the metric (\ref{Metric}). But, according to our understanding, those conditions 
are not necessary conditions for generating hair in the context studied here. Moreover, as we will see, the 
metric (\ref{Metric}) is a wormhole, not a black hole solution.

The radial electric field, from Eqs. (\ref{E}) and (\ref{E1}), is now written as
\begin{equation}
E(r)=\frac{Q}{r^2} \left( 1- \frac{8Mb^2}{r^3}+\frac{11b^2Q^2}{r^4}\right).
\label{BP}
\end{equation}
And, as expected, turning off $b$ the usual electric field is obtained. For large values of $r$,
the influence of $b$ on the electric field is very tiny. In this range of the radial coordinate,
the electric field is asymptotically the usual electric field for a point charge: $E(r) \rightarrow E_0(r)$.
  It is worth pointing out the weirdness of the new term that contains the mass
 parameter in Eq. (\ref{BP}), for it seems like the radial electric field was truly coupled to the spacetime geometry. 

As we said, the metric (\ref{Metric}) is an approximate solution with spherical 
symmetry in the Bopp--Podolsky electrodynamics.
We will see that the solution (\ref{Metric}) captures the same features pointed out in the 
 analysis carried out by Cuzinatto et al. \cite{Cuzinatto}, whose study is made
 from the Bekenstein method, even without a final expression for the spacetime metric. 
As we can read from Eq. (\ref{Metric}), the spacetime geometry is asymptotically flat, that is, 
the flatness condition 
\begin{equation}
\lim_{r\rightarrow \infty}A(r)=\lim_{r \rightarrow \infty}B(r)=1
\label{Asympt}
\end{equation}
 is achieved.
 Also it is worth emphasizing that the metric (\ref{Metric}) is solution of the Bopp--Podolsky 
 equation (\ref{Eq_Podolsky})
  in its approximate version, in which  high-order terms of $\epsilon$  were also ruled out.

\subsection{The spacetime structure}
By spacetime structure we mean the localization of horizons and, of course, type of infinities. 
As we saw in Eq. (\ref{Asympt}), the proposed
solution is, like the Reissner--Nordström spacetime, asymptotically flat, i.e., it has the same infinite configuration or
type of infinities of that well-known solution of Einstein's equations. 
But, supposedly, the metric (\ref{Metric}) presents a larger number of horizons. 
The localization of horizons in spacetimes like
(\ref{SS}) is given by the equation $g^{rr}=B(r)=0$ in cases when $A(r)$ and
$B(r)$ share the same roots. Typically, for $M>Q$,
the function $B(r)$ would have at most three positive and real roots, which would be horizons: 
two inner horizons (being the larger one
the usual inner horizon of the Reissner--Nordström spacetime, $r_-$) and 
one outer horizon, which is indicated as $r_+$ and would play
the role of the event horizon. However, as we said, due to the perturbative 
approach and the application of  spacetime (\ref{Metric}) to the shadow phenomenon, 
we are interested in the interval $r \geq r_+$, which is regular.

\begin{figure}
\begin{center}
\includegraphics[trim=0.4cm 0.4cm 0.6cm 0cm, clip=true,scale=0.6]{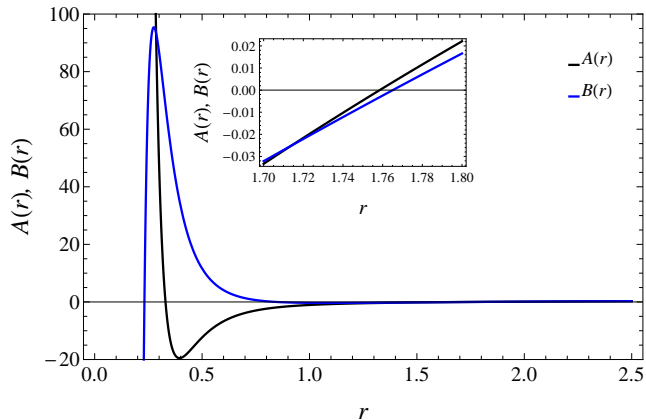}
\caption{The metric functions $A(r)$ and $B(r)$ given by Eq. (\ref{Metric}). The asymptotic behavior is indicated.
Also, one sees that the functions $A(r)$ and $B(r)$ present different zeros. This is one of the reasons why we interpret
such a solution as a wormhole solution.
In this graphic, one adopts $M=1,Q=0.7$ and $b=0.7$.}
\label{AB_graphic}
\end{center}
\end{figure}

However, the condition $A(r)\neq B(r)$ could raise questions about the metric signature and the 
existence of an event horizon.
As we can read from Eq. (\ref{Metric}), we adopt the metric signature $(+,-,-,-)$ , also
called Lorentzian signature. But from values of $r \leq r_+$, owing to the fact that $A(r)\neq B(r)$, 
we would have a non-Lorentzian signature because the outermost root of $A(r)$, namely $r_0$, would not coincide with
$r_+$ (see Fig. \ref{AB_graphic}), that is, $A(r)$ and $B(r)$ do not share the same root. 
Even worse, we would not have an event horizon, for the 
surface $r=r_+$ would not be a null surface. Then the line element $ds$, 
for constant $r,\theta$ and $\phi$, would not be zero at $r_+$. Here, we see a limitation of the chart $(t,r,\theta,\phi)$ 
for describing our spacetime.
  
In order to overcome this situation, we interpret the solution (\ref{Metric}) as a wormhole solution. 
Then the radial coordinate is valid just for $r>r_+=r_{\text{thr}}$. 
Now $r=r_{\text{thr}}$ indicates the wormhole throat localization in spacetime. 
Thus, instead of the event horizon, 
$g^{rr}=B(r)=0$ provides  an extremun of the radial
coordinate, $r=r_{\text{thr}}$. 
Consequently, the functions $A(r)$ and $B(r)$ are positive-definite functions for $r>r_{\text{thr}}$, and both
the metric is Lorentzian and the chart $(t,r,\theta,\phi)$ is still valid for $r_{\text{thr}}<r<\infty$. 
Following Refs.\cite{Molina,Molina:2012ay,Morris:1988cz},
 one defines then the proper length $\ell(r)$ as the new radial function, i.e.,
\begin{equation}
\frac{d\ell (r)}{dr}=\frac{1}{\sqrt{B(r)}}.
\label{ell}
\end{equation}
With the new coordinate $\ell (r)$, by assuming an appropriate
integration constant in Eq. (\ref{ell}), the region  $r_{\text{thr}}<r<\infty$ is mapped into $0<\ell < \infty$, and
 an analytic extension beyond $r=r_{\text{thr}}$ is viable. Thus,
we have two distant regions being connected by the wormhole hole throat at $\ell=0$. It is worth pointing out that the 
wormhole solution proposed here is traversable, that is, it has no event horizon. 
Moreover, the spacetime metric after the analytic extension $-\infty<\ell<\infty$ is regular, because from 
the Kretschmann scalar calculation the metric (\ref{Metric}) shows divergence at $A(r_0)=0$, in which
 $r_0<r_{\text{thr}}$. That is, $r=r_0$ is not part of spacetime anymore, thus the metric is regular. 

As we will see in the next subsection, the interpretation of (\ref{Metric}) as a wormhole solution is in agreement with the 
violation of the NEC and WEC, as several studies found out in wormhole
geometries.

\subsection{Energy conditions}
As mentioned in Cuzinatto et al. \cite{Cuzinatto}, where an analysis in the context adopted here 
was made without a final form for the spacetime metric, 
a spherically symmetric solution in the Bopp--Podolsky electrodynamics 
is able to violate energy conditions. In particular, as we said, both the NEC and the WEC are
 violated.  
From the total energy-momentum tensor given by $T_{\mu\nu}=T_{\mu\nu}^M+T_{\mu\nu}^a+T_{\mu\nu}^b$,
one has
\begin{equation}
T _{\ \nu}^{\mu}= \left(\begin{array}{cccc}
\rho(r)\\
 &-p_1 (r)\\
 &  & -p_2 (r)\\
 &  &  & -p_3(r)
\end{array}\right).
\label{Energy-momentum}
\end{equation}
For an energy-momentum tensor written as the diagonal matrix like Eq. (\ref{Energy-momentum}), the WEC states 
that $\rho(r)>0$ and, at the same time, the condition  $\rho+p_i\geq 0$ (for every $i=1,2,3$) should also be satisfied.
That is not the case for the metric (\ref{Metric}). As for the energy density, 
for the metric (\ref{Metric}) in the Bopp--Podolsky electrodynamics, we have
\begin{equation}
\rho (r)=\frac{Q^2}{8\pi r^4}\left(1-\frac{12b^2 }{r^2}A_0(r) \right) >0.
\label{rho}
\end{equation}
From the fact that $A_0(r)>0$ for $r>r_{\text{thr}}$, being the second term on the right side of Eq. (\ref{rho}) $\ll 1$,
 we can read that the energy density $\rho(r)$ is positive definite surely for $r > r_{\text{thr}}$, 
 which is the range of validity of the perturbative approach adopted by us.
 
On the other hand, as the condition $\rho+p_i\geq 0$ (for every $i=1,2,3$) is a true statement of either
energy conditions, thus both the NEC and WEC are not satisfied. In particular, for $i=1$, this condition reads
\begin{equation}
\rho(r)+p_1(r)= -\frac{3b^2Q^2}{\pi r^6}A_0(r),
\end{equation}
which vanishes for $b=0$ and is negative for $r>r_{\text{thr}}$, that is to say, the wormhole obtained here is not
in agreement with the NEC and WEC. Therefore, as a possible interpretation, 
 the model introduced in Section \ref{S2} could be translated into an exotic matter 
that does not satisfy energy conditions and provides a wormhole throat. 
In nonlinear electrodynamics, for example, according to 
Ref. \cite{Shabad:2011hf},
violations of energy conditions could lead to tachyons or signals faster than light in vacuum.

It is worth pointing out an interesting result in Ref. \cite{Cuzinatto} concerning 
the energy-momentum tensor. According to the mentioned article, the trace of  $T_{\mu\nu}$, which is
associated with the total energy of the system, would be finite for $r=r_{\text{thr}}$ independently of $b$. 
As the trace of the energy-momentum tensor is $T^{\mu}_{\mu}=T\sim 1/A(r)^2$, 
one has a finite value of $T$ when $r=r_{\text{thr}}$.

\section{The wormhole shadow}

\subsection{Radius of the shadow}
As the wormhole geometry (\ref{Metric}) has spherical symmetry, the shadow silhouette is circular. 
In general, shadows are deformed by the black hole or wormhole spin, which is not our case here.
According to EHT \cite{EventHorizonTelescope:2019dse,EventHorizonTelescope:2019ggy}, 
the M87* shadow presents
a small deviation from the perfect circle, a deviation from circularity $\Delta C \lesssim 10 \%$, indicating then
that M87* is spinning. But due to the less precise data from the Sgr A* image (because of intense variability of its
surroundings and the sparse coverage of the observations in 2017 \cite{EventHorizonTelescope:2022xqj}), 
the most precise parameter to be studied is the shadow angular size or angular diameter $d_{\text{sh}}$.

Black hole and wormhole shadows are generated by the so-called photon sphere, 
which is a region delimited by unstable photon orbits.
Part of photons traveling in these orbits falls into the hole, and part of them could escape to an observer like us.
For supermassive objects like M87* and Sgr A*, the shadow, a brightness depression, is surrounded by
a bright and thick ring of emission. According to EHT \cite{EventHorizonTelescope:2022xqj}, 
under certain circumstances, the size of the emission ring could be an approximate value for the shadow.  

\begin{figure}
\begin{center}
\includegraphics[trim=0.6cm 0.4cm 1.2cm 0cm, clip=true,scale=0.59]{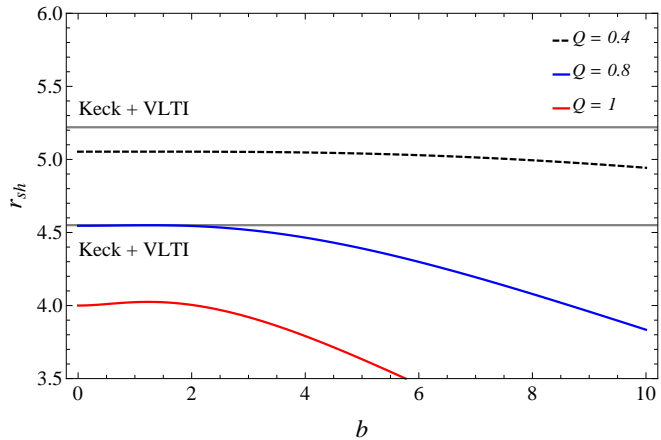}
\caption{The dependence of the shadow radius $r_{\text{sh}}$ on both 
the charge parameter $Q$ and coupling constant $b$. 
Exaggerated values of $b$ were chosen in order to better illustrate their effects on the shadow radius.
The parameter $b$ decreases the shadow radius.
The horizontal lines indicate the viable $1\sigma$ interval of the shadow radius, given by Eq. (\ref{Range}),
in agreement with the Sgr A* shadow radius using the mass-to-distance ratio from Keck + VLTI.
In this graphic, we adopt the geometrized units and $M=1$.}
\label{Shadow}
\end{center}
\end{figure}

The shadow silhouette is given by the photon sphere, and, according to Perlick and Tsupko \cite{Perlick}, 
from an interesting geometric argumentation, 
metrics like (\ref{SS}) present a photon sphere whose radius is given
by the following equation:
\begin{equation}
A(r_{\text{ph}})-\frac{1}{2}r_{\text{ph}}A'(r_{\text{ph}})=0,
\end{equation}
where $r_{\text{ph}}$ is the photon sphere radius. Following the mentioned article,
then the shadow radius $r_{\text{sh}}$ reads 
\begin{equation}
r_{\text{sh}}=\frac{r_{\text{ph}}}{\sqrt{A(r_{\text{ph}})}}.
\end{equation}
As expected, the shadow radius is not equal to the photon sphere radius. The gravitational lensing increases the shadow
radius when it is observed from a distance.
 Due to the high-order polynomial expression for calculating $r_{\text{ph}}$, we will omit the analytic result here. 
But for our purpose, the main information comes from the shadow radius, where its dependence on the
coupling constant is shown in Fig. \ref{Shadow}. As we can read 
from that figure, the coupling constant decreases the shadow radius. 
This could be a point in favor of our metric, for the EHT image disfavors alternative
black holes and wormholes that generate larger shadows than the Schwarzschild black hole.
However, as the EHT points out \cite{EventHorizonTelescope:2022xqj}, 
the most favorite candidate for the Sgr A* image is the Kerr metric, that is,
the observed image size is within $\sim 10 \%$ of the Kerr geometry predictions.

\begin{table}
\caption{Mass, in solar masses, and distance, in parsec, of Sgr A* according to Keck observatory and VLTI.}
\label{mass-to-distance}
\begin{ruledtabular}
\begin{tabular}{lcc}
     
      Survey & $M$ ($10^6 M_{\odot}$)& $D$(kpc)  \\ \hline
      \text{Keck\footnote{See Ref. \cite{Do:2019txf}.}} & $3.975 \pm 0.058$ &  $7.959 \pm 0.059$ \\ 
            \text{VLTI\footnote{See Ref. \cite{GRAVITY:2021xju}.}}&
       $4.297 \pm 0.012$   & $8.277 \pm 0.009$ \\
                         
\end{tabular}
\end{ruledtabular}
\end{table}

\subsection{Constraining the wormhole parameters using the Sgr A* shadow}
As we said in Introduction, in order to compare the geometry (\ref{Metric}) to the 
Sgr A* black hole, the crucial parameter here  is the deviation from the Schwarzschild metric $\delta$. 
As the Sgr A* spin is still unknown, geometries with spherical symmetry are not ruled out for describing
our galactic central compact object. A recent estimation for the dimensionless spin parameter for Sgr A*
is $a_{*} \lesssim 0.1$ \cite{Fragione:2020khu,Fragione:2022oau}, where $0 < a_{*} < 1$ 
for spinning black holes/wormholes, that is to say, arguably our central
object is spinning slowly. Thus, describing Sgr A* approximately 
as a static black hole (like the Schwarzschild black hole or
a deviation from the Schwarzschild geometry as a wormhole) is not out of question. 
Therefore, according to EHT \cite{EventHorizonTelescope:2022xqj}, 
the deviation from the Schwarzschild metric is conceived of as
\begin{equation}
\delta= \frac{d_{\text{sh}}}{d_{\text{Sch}}}-1,
\label{Deviation}
\end{equation} 
where $d_{\text{sh}}=48.7\pm 7.0$ $\mu$as is the shadow angular diameter of 
Sgr A* (measured by the EHT),
$d_{\text{Sch}}=6\sqrt{3}\theta_g$ is the angular diameter for the Schwarzschild black hole in the small angle
approximation, and $\theta_g$ is called angular gravitational radius, which is given by $\theta_g = GM/Dc^2$ 
where $M$ is the black hole or wormhole mass, 
and $D$ is its distance from us ($G$ and $c$ are the gravitational constant
and the speed of light in vacuum, which were set to 1 in the geometrized unit system adopted in this article). 
The most precise values for mass
and distance of Sgr A*, even adopted by the EHT, were obtained by the Keck observatory and
VLTI (Very Large Telescope Interferometer) (see Table \ref{mass-to-distance}). 
Both values provide the mass-to-distance ratio of our central
black hole/wormhole.
With those parameters for Sgr A*, the deviation given by Eq. (\ref{Deviation})
 is $\delta = -0.04^{+0.09}_{-0.10}$
for the Keck observatory, and $\delta = -0.08^{+0.09}_{-0.09}$ for the VLTI.

Following Vagnozzi et al. \cite{Vagnozzi}, the idea is calculating a valid range for the shadow radius of a 
wormhole like our spacetime (\ref{Metric}) using the deviation $\delta$. Spherically symmetric solutions---like the
Reissner--Nordström black hole, regular black holes, naked singularities or wormholes and black holes 
in several contexts like
 $f(R)$ gravity, braneworld, Horndeski model, bumblebee gravity, loop quantum gravity---were constrained
  by the authors.
For that purpose, Vagnozzi et al. \cite{Vagnozzi} obtained a combined value of $\delta$
using the two surveys, Keck and VLTI, considering them independent results. That procedure gives us
\begin{equation}
\delta \simeq -0.060 \pm 0.065,
\end{equation}
which leads to the  following $1\sigma$ constraint for the shadow radius as a deviation from the
Schwarzschild metric:
\begin{align}
3\sqrt{3}\left(1-0.125 \right) & \lesssim \frac{r_{\text{sh}}}{M} \lesssim 3\sqrt{3}\left(1 + 0.005 \right), \nonumber \\
4.55M &  \lesssim  r_{\text{sh}} \lesssim 5.22M.
\label{Range}
\end{align}
It is worth emphasizing that $r_{\text{sh}}$ is the shadow radius, not the angular radius.
Having said that, the range above constrains alternative geometries (like ours) allowed by the observations.
In this regard, Fig. \ref{Shadow} indicates upper bounds on the charge 
parameter, $Q \lesssim 0.8M$, and on the coupling constant, namely $b \lesssim 2M$.
We conclude then that our solution is still viable, for it was designed for $b\ll 2M$ according to the perturbative
approach.

\section{Final comments}
In this article, we built an approximate wormhole solution with spherical symmetry in
Bopp--Podolsky electrodynamics. The final form of the spacetime metric shows a third parameter, alongside
the mass and charge parameters, which comes from the action of the model and is conceived of as the
nonminimal coupling constant. Because of this parameter, the modified radial electric field violates massively
both the NEC and WEC at the wormhole throat.

In the final part of this work, data from the EHT Collaboration and from the Keck and VLTI observatories
were adopted in order to compare the wormhole obtained here with Sgr A*, the central compact object
in the Milky Way galaxy. These data constrain the range of validity of the shadow radius, 
when Sgr A* is considered as a small deviation from the Schwarzschild black hole. The new parameter
in the metric, the nonminimal coupling constant, which was constrained from the Sgr A* shadow,
makes the radius of the wormhole shadow smaller when it is compared with the Schwarzschild shadow, something
in agreement with the EHT report. 
As we pointed out, the metric obtained is viable using the 
valid range of the shadow radius of Sgr A*.
In this sense, Sgr A* could still be described as a wormhole.

\begin{acknowledgments}
LGM acknowledges CNPq-Brazil (Grant No. 308380/2019-3) for the partial financial support.  
JCSN thanks the ICT-UNIFAL for the kind hospitality. We thank two anonymous referees for comments and
valuable suggestions.

\end{acknowledgments}

\end{document}